\documentclass[aip,rsi,reprint,graphicx]{revtex4-1} 

\usepackage{graphics}

\draft 

\begin{document}


\title{Artifact-free data recovery system for an ASNOM application} 



\author{Dmitry Kazantsev}
\email[]{kaza@itep.ru}
 \altaffiliation{Perm.pos.: Institute for Theoretical and Experimental Physics, B.Cheremushkinskaya str. 25, 117218 Moscow}
 \affiliation{University of Erlangen-Nuremberg,
Cauerstr. 6, 91058 Erlangen}


\date{\today}

\begin{abstract}
 A digital signal acquisition system for an Apertureless SNOM (ASNOM)
based on a digital signal processing (DSP) card is presented. An
electromagnetic wave scattered by an AFM-like tip is initially
detected by an optical homodyning in a Michelson interferometer,
with a homodyne phase modulation caused by a periodic travel of the
reference arm mirror. Utilizing a non-linear dependency of the tip
scattering amplitude on a tip-sample separation distance, the
non-fundamental components of a tip oscillation frequency are
recovered in optical signal within each tip oscillation period. A
detected value, corresponding mostly to the near-field modes of a
tip-sample electromagnetic interaction, is then averaged over all
reference phases. A modular structure of the data acquisition and
signal recovery program allows modify easily the modes of operation
and the recovery algorithms.
\end{abstract}


%

\maketitle 

\section{Introduction}
An Apertureless Scanning Near-Field Microscopy\cite{s_SNOM_first}
(ASNOM) gets recently a powerful tool for the optical surface
investigations with a lateral resolution of a few
nanometer\cite{SNOM_1nm,Martin_ASNOM_Virus_APL1996}, regardless to
the wavelength being
used\cite{Knoll_APL1997,Keilmann_JM_2001_ASNOM_description}. The
successful applications of an ASNOM were demonstrated for
 a nanometer-scale material recognition\cite{Keilmann_Nat.399.134,Wurtz_ASNOM_RSI_98_1735},
 local field mapping\cite{Krenn_ASNOM_Plasmon_mapping_PRL1999},
 biological imaging\cite{Martin_ASNOM_Virus_APL1996,Keilmann_Virus_Map_NL2006}.
 ASNOM technique is also successfully implemented for a Raman
subwavelength resolution imaging
\cite{Anderson_TERS_APL2000,Hayazawa_TERS_AP2002}.

 A wavelength-independent nanometer-scale resolution can be achieved
due to the fact that an electromagnetic interaction between a
surface feature and a sharp AFM-like tip is caused by the evanescent
fields around the tip apex, which decay quickly with a distance.
 The oscillations of a tip dipolar momentum excited by a local
electromagnetic field around the tip are emitted into the
environmental space and can be reliably detected with some
experimental tricks.
 A sinusoidal tip oscillation used in a topography feedback system is
utilized in an ASNOM, to discriminate fundamental\cite{s_SNOM_first}
or, later, higher
\cite{Labardi_SecondHarm_ASNOM_APL2000,Keilmann_JM_2001_ASNOM_description}
harmonic components of a tip oscillation frequency $f_{tip}$ in an
optical signal.
 It is commonly accepted now, that a strong non-linearity of the
tip-surface near-field interaction dependency on the tip-surface
distance should be used to detect the higher harmonics $nf_{tip}$ in
a detector output, mainly consisting of an ASNOM signal.
 In first ASNOMs, the analog era single-frequency lock-in systems were
used\cite{SNOM_1nm,Maghelli_ASNOM_JM2001,Bek_ASNOM_RSI_2006_043703}
to detect the desired harmonic components. Last years, a wide
variety of the digital signal acquisition and processing systems got
available on the market, allowing more sophisticated recovery
algorithms. A digital recovery system was
proposed\cite{Ocelic_DSP_APL-2006} to detect a single high harmonic
component in a photocurrent.

A clear demand exists, to develop a flexible acquisition system for
the ASNOM applications, which can implement different modes of
operation: single-wavelength elastic light scattering, tip-enhanced
fluorescence microscopy, tip-enhanced Raman mode, "white light"
tip-enhanced Fourier-spectroscopy mode etc. Since an optical signal
in a fiber-defined SNOM also depends strongly on a distance between
the tip and surface feature, a signal recovery in such application
is very similar to one used in apertureless SNOMs. In some
applications, a gated photon-counting may be used instead of analog
signal processing.
 To achieve a flexibility, the data processing algorithms in such a
system must consist of independent software modules, so that any of
them can be easily improved or replaced without losses of a system
functionality.
 Just a few
papers\cite{Bek_ASNOM_RSI_2006_043703,Ocelic_DSP_APL-2006,esslinger_RSI_2012_033704}
are devoted to describe an ASNOM signal recovery.



\subsubsection{Hardware}
 In a present paper we report a successful attempt to use a DSP-based
signal acquisition card
P25M\cite{P25M_Manual_PDF_Web,P25M_Specs_Web} (Innovative
Integration Inc.). A card is equipped with four ADCs (max. rate
25~MSPS) and four DAC channes (max. rate 50~MSPS), 16-bit resolution
each. Parallel ports for a digital input and output are also
available on the card. Input-output operations (DACs, ADCs, digital
ports) are hardware-controlled with an FPGA logic, and the
parameters of its operation (clock frequencies, sampling frame
dimensions etc.) can be set by the software. A 300-MHz floating
point DSP (TMS320C6713) manages data acquisition, transfer and
output.

 A signal processing in the recovery system being described herein,
occurs in two hardware and software spaces. An input signal sampling
as well as analog output control are managed in a hardware/software
of the P25M card (further mentioned as \emph{target}).
 Math operations over the data having been collected, as well as
program-to-user interface, including a parameter control panels and
virtual oscilloscopes for a signal visualization are running in an
IBM-PC computer (which is mentioned as \emph{host} in the further
text). Finally, the recovered signal value can be sent from a host
computer to a target card in order to set appropriate voltages for
an analog transfer of a collected signal. For a data packet
transfer, a PCI slot on the host motherboard is used, in which a
target card is physically plugged in.

 For an SPM operation a R9 model SPM controller (RHK, Inc.) was used
and a home-made ASNOM scanning head\cite{Kaza_ASNOM_Head_Mi2013}.
Most of the measurements were carried out with a tunable $^{13}C
^{16}O_2$ laser (MERIT-G, Soliton Inc.) running in a CW mode
$(\lambda =10.6..11.3 \mu m)$.

\subsubsection{ASNOM signal: optical homodyning}
\begin{figure}
    \resizebox{0.45\textwidth}{!}{
    \includegraphics{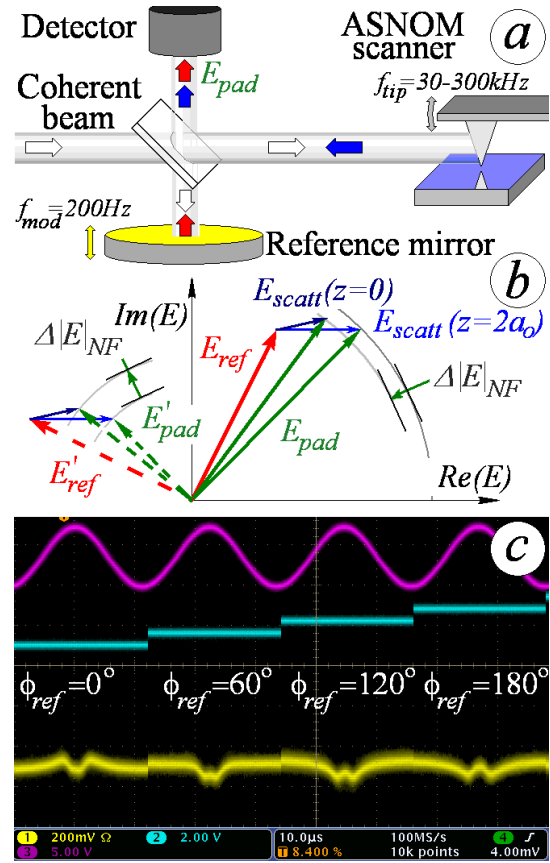}
    }
 \caption{
\label{E_Field_Figure}
 (a) Principle of a homodyne recovery of the optical signal in an ASNOM.
(b) Electromagnetic field amplitudes $E_{pad}=E_{scatt}+E_{ref}$ at
the sensitive pad of the photodetector, shown on a complex plane.
The amplitude/phase variations of the tip scattered wave (small blue
arrows $E_{scatt}$) are caused by the tip oscillation (frequency
$f_{tip}=30-300kHz$). A near-field optical interaction which occurs
between tip and surface at the moment of their "contact" modifies a
tip polarizability and therefore modulates $E_{scatt}$. Large (red)
arrows represent a field of the reference wave of nearly constant
amplitude $E_{ref}$ which phase is modulated by the periodic motion
($f_{mod}=100-200Hz$) of the reference beam mirror. Depending on the
reference wave phase, the detector response caused by the tip
oscillation can be positive or negative, as shown in (c): a tip
oscillation (top trace), and detector output signal (bottom trace)
at different reference phases (middle trace). A tip-surface
"contact" corresponds to the top of the sinusoidal tip deflection
trace.
}%
 \end{figure}

There are two key features in a signal recovery of an ASNOM
instrument running in elastic mode (in which the wavelength of a
scattered light does not change). First, a vanishing light wave
scattered by the tip under an influence of an irradiating field can
be "amplified" on a photodetector sensitive pad by optical
homodyning or heterodyning (see Fig.\ref{E_Field_Figure}(a)).
 In the presence of a coherent reference wave, the variations in the
photocurrent being detected are proportional to the \emph{first}
power of the scattered wave phasor length (instead of \emph{square}
expected without reference wave).
 Second feature utilizes a non-linear dependency of the tip scattered
wave amplitude on the tip-sample distance. Since it is the main
nonlinearity in the system, the higher harmonic components
$nf_{tip}$ in the photocurrent contain mainly the information
concerning an electromagnetic interaction between tip and surface,
caused by the fields localized in the gap between the tip and
sample.

 As it is shown in Fig.\ref{E_Field_Figure}(b), a photocurrent signal
in an optical homodyne Michelson scheme is determined by a sum of
the field phasors. Depending on a reference beam phase, such
interference can be constructive or destructive, so that a
photocurrent signal may contain positive or negative peaks which
appear at the moment when the tip "bounces" the surface. The traces
of a detector output signal, acquired simultaneously with a tip
oscillation signal are shown in Fig.\ref{E_Field_Figure}(c) for
different reference phases.
 For each definite reference phase, a near field component can be
extracted from the data, and stored as some complex number,
representing the amplitude and phase of a feature.

 One should keep in mind, that recovery of just a single
non-fundamental harmonic (let say second) content is not perfectly
adequate to the signal being expected and collected. The
near-field-caused pulse in the signal has a Gaussian-like shape, as
shown in Fig.\ref{E_Field_Figure}(b). It means that all higher
harmonics contain the information together (and their phases must
all be acquired precisely), so that a recovery of just a single one
means some loss of a useful signal. More careful investigation of
this pulse-like photocurrent feature shown in the figure, leads to
an observation that not only amplitude of a scattered wave depends
on a tip-sample gap, but also a phase: the trace looks as if it was
plotted on a side of a cylinder, rotating with a reference mirror
displacement. Therefore, to improve a signal to noise ratio, some
gated recovery algorithm looks to be better, with a gate shape
corresponding to an expected signal shape. It seems useful to make a
remark here that a gated photon
counting\cite{Bek_ASNOM_RSI_2006_043703} can be implemented for a
signal recovery as well.

 Regardless to the near-field pulse recovery algorithm within a
single tip oscillation period, an averaging over all possible phases
of a reference beam must be provided afterwards, to exclude random
nature of a reference arm length.
 It is clear from the figure, that the
oscillations in a photocurrent signal, caused by near-field
interaction between tip and surface, must depend on the reference
phase in a sinusoidal way:
 \begin{equation}
 \Delta I_{det} \propto \Delta E_{scatt} E_{ref}cos(\phi _{ref})
 \label{eq:Delta_I_det}
 \end{equation}
 Thus, to get an adequate measure of a near-field tip-sample
interaction, an acquired signal must be averaged over all possible
reference beam phases. Most evident way\cite{Ocelic_DSP_APL-2006} is
to modulate an optical length of a reference arm by a periodic
mirror motion and to consider a phase modulation respectively in a
recovery math. Another way to average over all reference beam phase
is a heterodyne
recovery\cite{Hillenbrand_OptHeterodyne_PRL2000,Keilmann_JM_2001_ASNOM_description}.
In this case, a reference field phasor turns on a complex plane
around with a rate of a light modulation frequency ($\Omega
_{mod}=80~MHz$ in the papers being cited). A recovery of a signal at
this combinational frequency $\Omega _{mod}+nf_{tip}$ yields a
measure of an ASNOM signal, averaged over several turns of a
reference phase.

 In an assumption that there is completely no artifacts caused by
the optical misalignments, and a travel of a reference mirror leads
purely to an appropriate turn of the reference field phasor, a
reference phase modulation can be simplified by using of just two
points of a mirror
position\cite{Vogelgesang_Nanodisks_DoubleScan_NL2008}, shifted with
respect to each other for $90^{\circ}$ of the phase delay. It leads
to an obvious disadvantage: each raster line should be mechanically
scanned twice, with different reference phases. Additionally, the
operator looses a degree of freedom to discover a misalignment of an
interferometer (e.g. by observing a wrong shape of a signal trace on
the oscilloscope) before mapping got started.

Therefore, to detect an ASNOM optical signal two math procedures are
necessary. First, a non-fundamental response feature in the signal
must be recovered within each tip oscillation period, yielding in
some complex value. Such a processing we will mean as \emph{primary
math} in a further consideration. Second, the results of primary
math must be averaged over full turn of a reference beam phase
(keeping in mind the reference beam amplitude turn on a complex
plane). Such an operation will be mentioned as \emph{secondary math}
in the further text. It looks to be more-less natural to separate
all processing steps, so that any of them took a minimal influence
on the other flow stages.

 \begin{figure}
    \resizebox{0.45\textwidth}{!}{
    \includegraphics{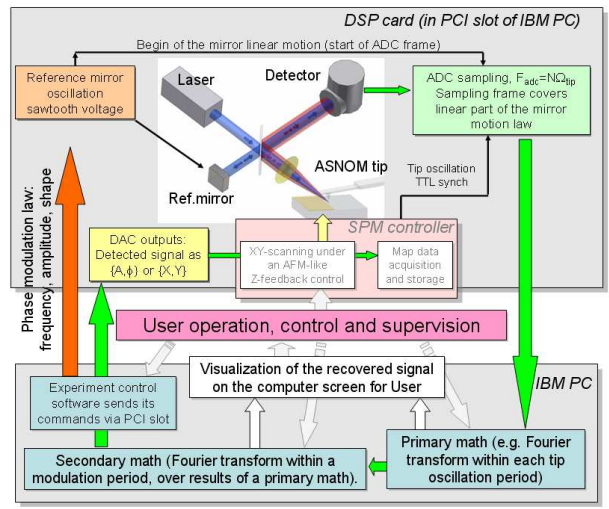}
    }
 \caption{
\label{Recov_Flow_Figure}
 Signal recovery flow in an ASNOM system. An optical signal from a
detector at the interferometer output is digitized by an ADC in the
acquisition board. Sampling frequency is set exactly to a multiple
($\times 32...\times 128$) of an ASNOM tip oscillation frequency,
providing enough resolution within an oscillation period and
constant acquisition phase over the whole sampling frame. A mirror
in a reference arm of an interferometer is driven along its axis
($f_{mod}\approx 100..200~Hz$) to provide a phase modulation in an
optical homodyning. An ADC sampling frame starts at the begin of a
linear part of the mirror working travel, and covers whole period of
this linear part. Non-fundamental components of the tip tapping
frequency in the photocurrent, caused mainly by a tip-sample
near-field interaction, are recovered within each period of the tip
oscillation (primary math). An yield of that (a complex number for
each tip oscillation period) is then used as input data for a
secondary math. An array of primary math results is utilized in a
Fourier-transform procedure to average the response over all
reference beam phases (secondary math). A result of that (a complex
number, corresponding to a near-field tip-sample interaction) is
sent to an SPM controller to be stored together with other map data
like topography.
}%
 \end{figure}

A signal recovery data flow is shown schematically in
Fig.\ref{Recov_Flow_Figure}.

 \section{Reference mirror motion control}
To provide a phase modulation, a reference mirror should be driven
along its axis. If its positioning is described with a sinusoidal
law, some unpleasant conditions appear in the game. A phase $\phi
_{ref}$ of a reference beam is determined by the arm optical length
$L$ in a following way: $\phi=2\pi \frac{2L}{c}=2\pi \frac{2L_0
}{c}cos(2 \pi f_{mod} t)$
 (a factor of two near variable $L$ represents that the light
propagates forward and back in the reference arm). In this case, a
dependency on time for the reference wave phasor in
 (\ref{eq:Delta_I_det})
 becomes double-sinusoidal:
 \begin{equation}
 \label{eq:E_cos_cos}
 E_{ref}cos(\phi _{ref})= E_{ref}cos(2\pi
 \frac{2L_0}{c}cos(2 \pi f_{mod} t))
\end{equation}

It means that, in the case of sinusoidal law of the phase
modulation, the data points provided by the primary math are
distributed non-uniformly on a phase scale. Their density is higher
near the dead points of reference mirror working stroke, and lower
in the middle. It makes an utilization of a standard
Fourier-transform operation problematic. Additionally, such a
non-uniform distribution may lead to appearance of artifacts, caused
by enhancement of the noise containing in some data group and
suppression of the signal in others. Another solution to overpass
that might be a variable ADC sampling rate within a phase modulation
period, but it is technically rather hard to manage.

Due to these reasons, a voltage which determines a position of the
reference mirror (being applied to the piezo-actuators) must contain
a linear part in its oscillation law. In a single-wavelength
application, a span of this linear part must correspond to half a
wavelength, so that a reference phase makes a full turn during the
data acquisition. In a multiple-wavelength applications, with a
signal recovery algorithm similar to the Fourier-spectroscopy, the
mirror working travel must be naturally much longer than a
wavelength used in the experiment.

The sharp shocks at the dead points, corresponding to a motion
direction change, may cause parasite axial and angular vibrations of
the reference mirror, and spoil an alignment of the interferometer
scheme. To avoid that, some smoothing of the oscillation law must be
provided for a vicinity of the dead points. Naturally, it enlarges
real range of motion, as well as decreases a fraction of time
available for the data acquisition. We use a semi-period of a
sinusoid, tailored by the derivative to the linear part of the
curve.

 An array of 16-bit data, corresponding to the desired oscillation
law (with appropriate values of amplitude, position offset, dead
point smoothing etc.) is transferred by a target CPU to the DAC
output queue in a target FPGA. The DAC clock frequency, physically
kept by an FPGA circuitry, is also loaded into an appropriate
control register of FPGA once at initialization time by the program
running in the target CPU. Thus, a signal of appropriate shape and
frequency is generated in DAC/FPGA with no further intervention of
the target CPU.

\begin{figure}
    \resizebox{0.45\textwidth}{!}{
    \includegraphics{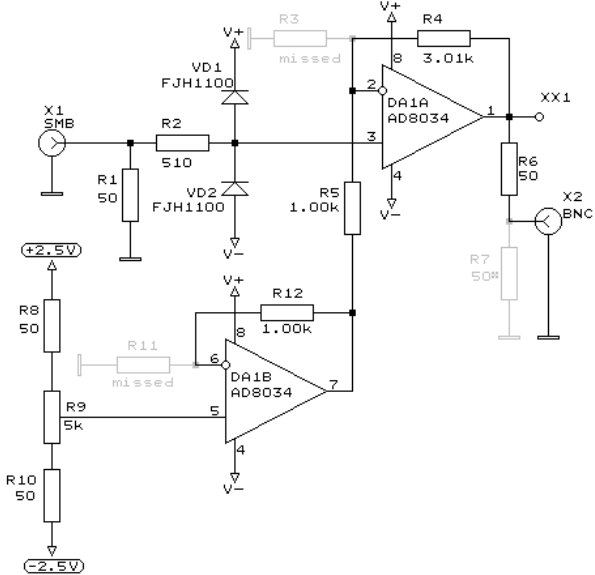}
    }
 \caption{
\label{DAC_Stage_Figure} A DAC output voltage level conversion
buffer stage. The voltage span of $(-1.25V..+1.25V)$ is converted to
a range of $(-1.0V..+9.0V)$ required for an input of the controller
which drives piezo stage where a reference mirror is mounted. An
input preamplifier board is similar, which converts the voltages to
a range of the ADC 50-Ohm inputs. The shadowed components are
mounted in that case.
}%
 \end{figure}

 To convert a voltage from (-2.0V...+2.0V) range provided by a P25M
card DAC output to a (-12V...+2V) range required by an input of the
controller \cite{E503_HV_Amplifier_Specs_Web}
which drives a S-316 piezo-positioning stage
 \cite{S316_Stage_Specs_Web,S316_Stage_Datasheet_Pdf_Web}
, a home-made amplifier board was used. It contains four independent
stages with adjustable gain and offset each
 (see Fig.\ref{DAC_Stage_Figure}).
 To reduce the noises, its power supply is isolated from all nets of
the host computer (IA0515D followed by LM7812/LM7912), and its
ground is connected just to output jack of the P25M card.

 Due to accepted ideology,
 a new value of a recovered signal
gets available once per a mirror oscillation period (half period, if
ascending and descending branches of the mirror oscillation law are
processed separately). Therefore, a reference mirror oscillation
frequency limits a final speed of an ASNOM mapping in the system,
and it seems natural to increase it as much as possible. There are,
however, two limiting factors. First, the oscillation period of the
ASNOM tip must be much shorter than a mirror oscillation time, in
order to simplify data processing by an assumption that the
reference phase was constant during a tip oscillation period. With a
typical resonant frequencies of the tips being used ($f
_{tip}=30..300~kHz$) such a condition seems to be fulfilled if
$f_{mod}$ is less than $1..3~kHz$. Second factor is a limited
mechanical response rate of the piezo-actuator being used for a
mirror positioning. A feedback-enhanced positioning stage
 \cite{S316_Stage_Datasheet_Pdf_Web}
S-316 (Physik Instrumente, GmbH), is claimed to have its first
resonance frequency of 5.5~kHz. Nevertheless, a small commercial
mirror (PF05-03-M01, Thorlabs) is already too heavy for this stage,
reducing the frequency of an artifact-free operation to approx.
120~Hz. Situation can be improved by using of a thick Au-coated
piece of a silicone wafer instead of a mirror, but even in that case
the shape of the acquired primary signal gets distorted at the
frequencies above 220~Hz, indicating most probably an angular
misalignment of the interferometer.

%
%


\section{Data acquisition}
 To acquire the data, an ADC channel of the signal acquisition board
is used. A sampling frame covers full linear part of the reference
mirror motion. As a hardware event to trigger a frame, the edge of
the tip oscillation clock (received as a TTL signal from an SPM
controller) is used, which comes first after a linear part of the
reference mirror is started (software event generated by the mirror
control thread in the target space). A sampling rate is set to an
exact multiple of $f _{tip}$, so that each tip oscillation period is
covered by 32..128 samples of ADC. Thus, there should be enough data
to restore higher harmonic components in the signal.

 To avoid an aliasing and/or phase unambiguity in the received data
during the whole reference phase turn, an ADC sampling frequency is
set to exact multiple of the tip oscillation frequency.
 The acquisition frame parameters (ADC sampling frequency, logical
length of a sampling frame etc.) are stored into the FPGA registers
on a target bus once at the initialization time. To be triggered, an
acquisition of a new frame requires just a few CPU operations, and
produces therefore almost no operational delay in the system.

Another ADC channel is simultaneously used for a sampling of a tip
displacement signal, received from an ASNOM scanning head.
Nominally, an AFM-like feedback of the SPM controller keeps the
conditions of a tip oscillation constant. In a real life, the phase
of a tip oscillation can be slightly shifted from its ideal value.
It means, that a recovery of an ASNOM pulse in the photocurrent
(which occurs just at the moment of the tip-sample "contact") can be
distorted by a time shift between ideal and real moment of a
contact. The system would be especially sensitive to such a time lag
in the mode of gated recovery of the signal (which is more promising
than a single-harmonic analysis, because it utilizes a whole
spectrum of an ASNOM response). The ADC samplings of a tip
displacement signal, being controlled by an FPGA circuitry, occur
exactly simultaneously with the samplings of an optical signal
delivered from a detector.

As soon as data frame is completely acquired in the target, it is
sent to the host, via their common PCI-channel.

The third ADC channel of a target card can be used for a control of
voltage driving the reference mirror. It allows visualize all
relevant processes on the screen of the same virtual oscilloscope in
the host computer. It should be kept in mind, however, that a
transfer of data acquired also by the second pair of ADC inputs
reduces a rate of a PCI channel being used for a communication
between a target and a host.

 \begin{figure}
    \resizebox{0.45\textwidth}{!}{
    \includegraphics{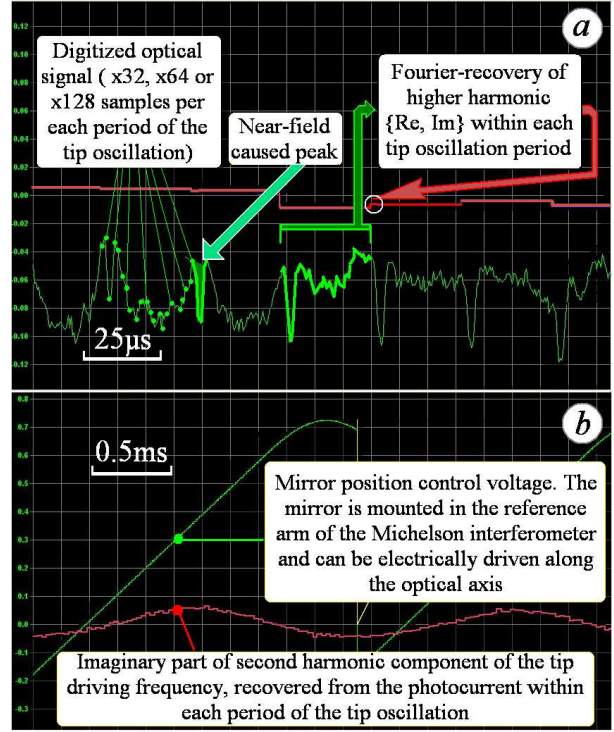}
    }
 \caption{
\label{Signal_Processing_Figure}
 A screenshot of the virtual oscilloscope included in a signal recovery
program. (a) Primary math. The optical signal data points ($\times
32...\times 128$ per each tip oscillation period) are first
processed to extract a complex number representing a near-field
component within each period separately. An imaginary part of result
is plotted as a red trace, a real one (not shown) looks similar. (b)
Secondary math. The results of a primary math (its imaginary part is
shown as a red trace) are used as an input for a Fourier-transform,
which takes into account a reference beam phase sweep (see a
ramp-like trace, stored simultaneously with other input data).
Finally a result of this procedure is sent to a SPM controller as a
recovered near-field signal.
}%
 \end{figure}

\section{ASNOM component recovery within a tip oscillation period (primary math)}
 After a host receives a new data frame, a recovery of the peaks
caused by a near-field effects gets started. A full data frame,
acquired for a whole working travel of the reference beam mirror is
subdivided before processing into the logical pieces corresponding
to a single tip oscillation period each. Since an ADC sampling
frequency is set to exactly multiple of a tip oscillation frequency
$f_{tip}$, all periods in the data array are acquired at completely
same phase, so that the jitter and aliasing artifacts are
suppressed.
 There is a choice of a recovery algorithms to be used at this step
of signal discrimination. First, just a single higher harmonic
component can be extracted as a complex number within a period.
Second, a detector signal can be multiplied by a gate function,
corresponding in time to the moment of a tip-surface touch. Third,
the pulses received from a photon counter (in that case a frame of
digital port data must be acquired from a target) can also be used
as an input data.

Simultaneously to the optical signal recovery, a first harmonic
component is calculated in a data set corresponding to a cantilever
deflection signal. Ideally, a topography feedback should keep the
conditions of a tip oscillation to be constant. In reality, the
phase of a deflection suffers of significant fluctuations ($\pm
10..20 ^{\circ}$), and therefore an exact time of a "bounce" can be
shifted. A measured phase shift of a real tip oscillation law allows
to correct a position of a gate function in a gated discrimination,
or to correct a phase of output data for a higher harmonic
demodulation method.

\section{Averaging over different phases of a reference beam (secondary math)}

Since a reference modulation law was chosen to be linear in a time,
an average over the reference beam phase consists simply of a single
harmonic calculation in an array of complex numbers containing the
results of a primary math. In a simplest case of elastic scattering
of the light, a linear part of the mirror oscillation law is set to
cover a bit more than a half wavelentgh (full turn of a reference
phase), so that a primary math data contain just a single
oscillation of a value. Nevertheless, there is no principal
limitations to deal with a multiple-period data (in such a case some
higher harmonic should be calculated) or even to do full
Fourier-transform processing like in a Foruer-spectrometer if the
reference mirror covers much more than a single wavelentgh.

Since the reference phasor rotation is opposite for an ascending and
descending branches of a reference mirror travel, we subdivide data
processing for these branches in two separate jobs. While a target
acquires a data for an ascending branch, a host processes the data
of descending branch, received before. Once ascending branch data is
sent to the host, it will be processed in a turn.
 The result of a data processing is shown in a virtual oscilloscope
on a host screen, so that an operator can control the amplitude,
phase and shape of the signal and fix the artifacts if they appear
in the experiment. In particular, the operator can decide which
correction phase should be used before the ascending and descending
branch results are added together. At the high phase modulation
frequencies, non-ideality of the mirror positioner is noticeable, so
that the real reference phase is not straight determined by the
control voltage being applied from the DAC.

 Generally it means that a final data appears at the system output
(in a digital form to be sent into a SPM controller via software or
as analog voltage output of DAC) always with a delay of a phase
modulation period.

\section{Synchronization}
 The signal acquisition, transfer and recovery processes, as well as
operation control and other processes of code/data preparation and
transfer, are programmed as separate threads "simultaneously"
running in host and target.
 A data acquisition thread in the target, which starts an ADC
sampling frame and transfers the packet to the host once the data is
acquired, stays inactive until the reference mirror control thread
puts a command in a mailbox after it has loaded a new portion of an
oscillation law data in a queue of DAC output. A mirror oscillation
control thread in the target wakes up on a hardware interrupt, which
occurs on an underflow of a DAC data queue in the FPGA logic.
Started on this event, the mirror control thread fills the output
buffer of DAC in the FPGA, reports a beginning of a new oscillation
period to the ADC acquisition control thread and gets asleep again.

In less requiring cases, the control over the thread operation (e.g.
change of parameters for the next operational cycles) is provided by
the function calls. In these cases, as a rule, the new parameters
are not implemented immediately. Instead, they get stored in some
variables of the thread classes. An execute loop of the thread
checks before the operational cycle start if new parameters were
received and then initializes appropriate hardware/software
 if necessary.

 Some software operations require even more time than a typical
working cycle of the thread. In such a case, the operation (e.g.
disk output of data) is entrusted to another thread of low priority,
and a calling thread can continue its main job.
 To generate a data array corresponding to a reference mirror
oscillation law (with its dimension corresponding to the DAC clock
frequency, and data content corresponding to a voltage being applied
at output), a low-priority thread runs simultaneously with a DAC
control thread by its order. Thus, an output voltage does not suffer
of a voltage freezing if the parameters of a modulation law are
modified. After new oscillation law data is ready, the data
preparation thread informs a DAC control thread, and the law is
replaced in its data table.

 Similar things occur in the host software. A data recovery thread
gets active on a receiving new frame of and DAC samples (from a
packet transfer dispatch class). Once the calculations (primary and
secondary math) are done, it sends the result (a pair of numbers,
representing a complex number acquired for the last mirror
oscillation period) into the dispatch class to be transferred into
the target DACs for an analog output.

To prevent the PCI bus jams, the host sends a confirmation of each
successful input data frame processing to the target. If, by some
reasons, the rate of a host software was not enough to complete a
previous data (e.g. an operational system has blocked the activity
of a signal processing program), the target does not start a new
acquisition followed by a PCI-transfer of a data frame.

\section{Data output}

 The results of an ASNOM-relevant signal recovery are sent into the
SPM controller after each period of a reference mirror oscillation.
Generally, each result sample is a complex number, and therefore two
floating point values should be transferred (or two integers
representing them). For an analog transfer, two DAC channels of a
target card can be used, and their outputs should be connected to
the appropriate analog inputs of an SPM controller. Also, the data
can be transferred to the SPM control software via IP connection or
a pipe object of a Windows-XP operating system.

 Also, an option to save any frame data to a hard disk is available.
All data received from a target (typically a frame duration
corresponds to a linear part of a reference mirror motion law) are
saved as an ASCII table or in a binary format, on a user command
"SaveFrame" issued by a mouse click on a GUI panel.
\section{Artifacts known}

A periodic axial displacement of a reference arm mirror, may produce
some artifacts. A position of a mirror which is mounted on a
piezo-stage may differ from a control voltage applied to the stage
controller, even despite an operation of the stage controller
feedback.

 \begin{figure}
    \resizebox{0.45\textwidth}{!}{
    \includegraphics{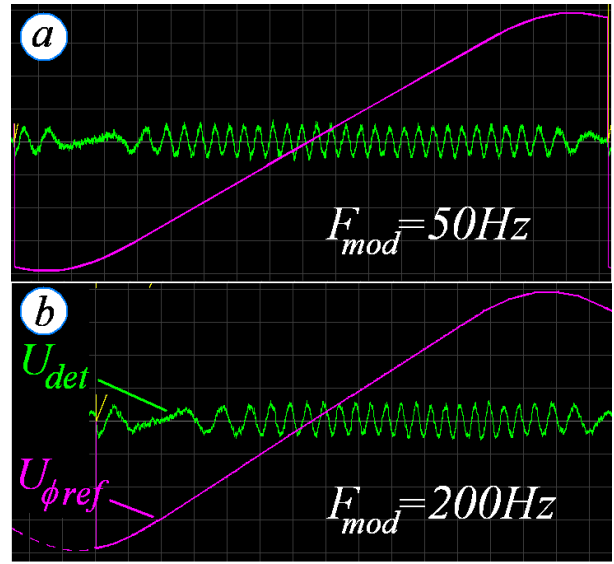}
    }
 \caption{
\label{Mirror_Nonlinearity_HeNe_Figure}  A non-linearity of the
reference mirror positioning. An objective focussing the signal beam
onto an ASNOM tip is replaced with a flat mirror, to get a purely
Michelson scheme. A HeNe laser (wavelength of 632~nm) is used
instead of $CO_2$ laser to increase a number of observable
interference periods. (a) phase modulation frequency is $50~Hz$, (b)
modulation frequency is $200~Hz$. A green trace represents a
detector output signal $U_{det}$, a violet trace represents a mirror
position control voltage $U_{{\phi}ref}$ acquired by another ADC
channel simultaneously. An interferometer output signal shows that a
real position of the reference mirror does not correspond perfectly
to a control voltage being applied. That fraction of period, in
which a reference phase depends on time in a linear way, shrinks at
the higher modulation frequencies for a thick (heavy) mirror being
used.
}%
 \end{figure}

 There is also an artifact found, specific for a primary math used
in a system being described. If a sample contains some large
light-scattering features (e.g. an edge of the sample wafer or a
metal structure on a semiconductor surface), a light wave scattered
by the feature back to the interferometer provides an interference
with a reference beam.

 \begin{figure}
    \resizebox{0.45\textwidth}{!}{
    \includegraphics{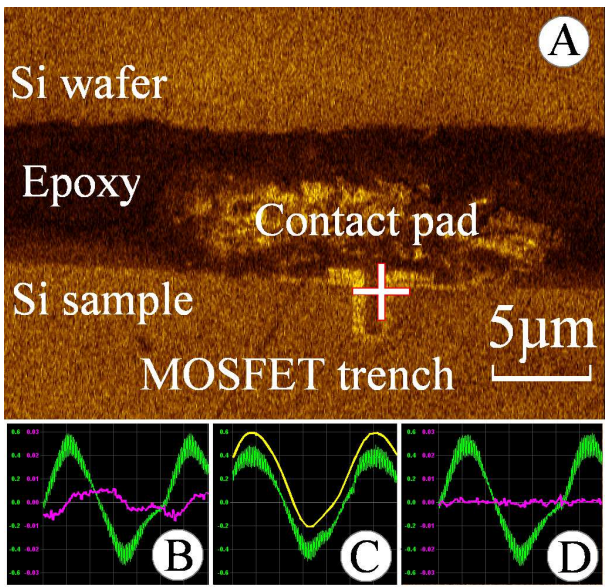}
    }
 \caption{
\label{SignalSlopeArtifact_Figure} (A) ASNOM map of the
trench-defined Si-MOSFET transistor structure (image bottom),
observed in a cleaved edge geometry.
 After a mapping, a tip was retracted from the
surface for about 2~${\mu}m$ at the location marked with a cross. No
ASNOM signal should be detected in such a mode.
 (B) Detector output signal (green trace).
 An interference of the light scattered by the sample structures and
cantilever with a reference beam causes nearly sinusoidal
oscillation in an average value of the signal, due to reference
mirror motion. This oscillation produces an average slope in the
input data for the primary math Fourier-transform within each
$f_{tip}$ period. A signal recovered at $2f_{tip}$ for a tip
oscillation period (violet trace) contains therefore a non-zero
component, modulated by the reference phase. Such a component
becomes a parasite signal for the further operations.
 (C) a curve of a smoothed signal (yellow trace) should be subtracted
from the input data (green trace) before primary math. (D) as a
result, no artifact is present in a recovered ASNOM signal (violet
trace).
}%
 \end{figure}

 As an example, an image of the Si trench-defined MOSFET
structure is present in the Fig.\ref{SignalSlopeArtifact_Figure}. A
sample was prepared in a cleaved-edge geometry. To protect the edge
during polishing, a piece of dummy Si wafer was glued on a sample
carrier parallel to a sample. An ASNOM signal amplitude mapped at
wavelength $\lambda=10.68{\mu}m$  is shown in
Fig.\ref{SignalSlopeArtifact_Figure}A.
A $2^{nd}$ harmonic of the tip oscillation frequency $f_{tip}$ was
used in the primary math to recover a near-field component. A trench
structure ($amSi$ surrounded with $SiO_2$) covered with some metal
 contact strip can be seen in the figure. After map was
obtained, a tip was brought to a position above the trench (marked
as a cross in the map) and retracted from a surface for approx.
$2{\mu}m$. Since a tip-surface distance is much larger than
necessary for a near-field optical coupling, one should expect
absolutely no near field component in the signal being observed.
Instead, one can clearly see nearly sinusoidal oscillation in a
second harmonic response (Fig.\ref{SignalSlopeArtifact_Figure}B,
violet curve) in an output data of the primary math.

 Due to a presence of the sharp edges and significant difference in the
optical properties of the materials ($Si$, $amSi$, $SiO_2$, $Al$,
epoxy), the light scattered by the sample back is rather strong, and
a parasite interference component is noticeable. To save a CPU time,
no data apodization is applied before Fourier-processing.
 In this case, if a Fourier-transform is used to recover an ASNOM
component in a detector signal within each tip oscillation period
(higher harmonics of $f_{tip}$), the slope and curvature of this
slow interference response would be processed as well as any other
input signal by the Fourier-transform math, and produce some
non-zero result (see Fig.\ref{SignalSlopeArtifact_Figure}B). The
higher modulation frequency is, the stronger is the effect. Such a
"recovered signal" has no physical reason in the terms of a
near-field interaction between the tip and surface, and should be
considered as a math artifact. To clean this parasite component
away, a smoothed curve of an input signal (see
Fig.\ref{SignalSlopeArtifact_Figure}C) should be subtracted from the
detector data before a primary math. Since just a general shape of a
parasite interference law is relevant for the correction, the
smoothening curve is built on the basis of signal average values
calculated within each $f_{tip}$ period. This (relatively short,
$N=\frac{f_{tip}}{2f_{mod}}=50..1000$) array of numbers is used to
build a fit curve. Just a few first (3..5) harmonic components
within this array are calculated to describe the curve shape. After
that, these constants are used to interpolate the correction curve
inside a period of the tip oscillation ($\times 32..\times 128$
points for each period, depending on ADC sampling rate being set).
To save a CPU time, Tailor series (with coefficients calculated from
the data mentioned above) could be used for interpolation instead of
Fourier series. Finally, a parasite component could be cleaned away
completely (see Fig.\ref{SignalSlopeArtifact_Figure}D), and this is
rather important for an artifact-free ASNOM imaging of the
weakly-scattering samples, which produce low useful signal.

\section{Conclusions}
 A sensitive, flexible data acquisition and signal recovery system is
developed. A modular structure of a software makes easy modify the
functionality of the system. A clear subdivision of the data
processing stages allows user visually control the intermediate data
on a virtual oscilloscope screen and detect the misalignments in the
system by the signal shape distortions being observed.

Even a modest CPU rate of 300~MHz in a data acquisition card and of
2.66~GHz (Quad-CPU) in a host computer is enough to recover an ASNOM
component in a detector response without losses of data.
Nevertheless, we should mention that an ADC sampling rate above
6~MSPS makes an operation less stable, so that the target card may
hang in some minutes or hours. It was found also, that a DAC
sampling rate above 100~kSPS may cause some troubles at a high ADC
sampling rates. Nevertheless, these sampling rates are much higher
than ones necessary for a successful operation. The most limiting
feature in our opinion is the mirror oscillation frequency, because
a single data point of an ASNOM map can be acquired just within of a
mirror oscillation period. With a modulation frequency of 200~Hz it
means that a single point acquisition requires 5~ms (or 2.5~ms if
ascening and descending mirror travels are sent to output
separately).

\section{Acknowledgements}
The work was supported by a German National Science Foundation (DFG
grant KA3105/1-1). Author thanks Dr. R.Hillenbrand, Dr. F.Keilmann
and Dr. N.Ocelic for their valuable discussions concerning
principles of an ASNOM signal recovery.


%


\bibliography{Kazantsev_ASNOM_DSP_Recovery}

\end{document}